\documentclass[iop]{emulateapj}
\slugcomment{Accepted to ApJ January 11, 2016}
\usepackage[varg]{txfonts}
\usepackage{microtype}

\newcommand{\sups}[1]{\ensuremath{^{\textrm{\scriptsize{#1}}}}} 
\newcommand{\subs}[1]{\ensuremath{_{\textrm{\scriptsize{#1}}}}} 

\newcommand{\tr}[1]{\ensuremath{\textrm{#1}}} 
\newcommand{\JM}{\ensuremath{M_J\,}}
\newcommand{\EM}{\ensuremath{M_\oplus\,}}
\newcommand{\sauthor}[2][]{\author{#2\sups{#1}}} 
\newcommand{\saffil}[2][]{\affil{\sups{#1}#2}}

\def\therefore{
\leavevmode
\lower0.1ex\hbox{$\cdot$}
\kern-0.1em\raise0.7ex\hbox{$\cdot$}
\kern-0.1em\lower0.1ex\hbox{$\cdot$}
\;} 

\usepackage{float}

\usepackage{natbib}
\usepackage{color}

\newcommand{\pic}{./}
\newcommand{\thebib}{../../tex_stuff}

\begin{document}

\title{Hiding in the Shadows \textrm{II}: Collisional Dust as Exoplanet Markers}
\sauthor[1,2]{Jack Dobinson}
\sauthor[1]{Zo\"{e} M. Leinhardt}
\sauthor[1]{Stefan Lines}
\sauthor[1]{Philip J. Carter}
\sauthor[3]{Sarah E. Dodson-Robinson}
\sauthor[2]{Nick A. Teanby}
\saffil[1]{University of Bristol, School of Physics, H. H. Wills Physics Laboratory, University of Bristol, Bristol, BS8 1TL, UK}
\saffil[2]{University of Bristol, School of Earth Sciences, H. H. Wills Physics Laboratory, University of Bristol, Bristol, BS8 1TL, UK}
\saffil[3]{University of Delaware, Department of Physics and Astronomy, 217 Sharp Lab, Newark, DE 19716, USA}



\begin{abstract}
Observations of the youngest planets ($\sim$1-10 Myr for a transitional disk) will increase the accuracy of our planet formation models. Unfortunately, observations of such planets are challenging and time-consuming to undertake even in ideal circumstances. Therefore, we propose the determination of a set of markers that can pre-select promising exoplanet-hosting candidate disks. To this end, N-body simulations were conducted to investigate the effect of an embedded Jupiter mass planet on the dynamics of the surrounding planetesimal disk and the resulting creation of second generation collisional dust. We use a new collision model that allows fragmentation and erosion of planetesimals, and dust-sized fragments are simulated in a post process step including non-gravitational forces due to stellar radiation and a gaseous protoplanetary disk. Synthetic images from our numerical simulations show a bright double ring at 850 $\mu$m for a low eccentricity planet, whereas a high eccentricity planet would produce a characteristic inner ring with asymmetries in the disk. In the presence of first generation primordial dust these markers would be difficult to detect far from the orbit of the embedded planet, but would be detectable inside a gap of planetary origin in a transitional disk.
\end{abstract}

\maketitle

\section{Introduction}

Over the last twenty years more than 1900 exoplanets have been discovered with a huge diversity in system parameters. These discoveries imply that planet formation is a ubiquitous phenomenon. In order to discriminate between different models of planet formation, observations of evolved planetary systems are of great utility. However due to the chaotic nature of planetary dynamics, many formation models produce end results that are indistinguishable. Observations of young exoplanets would discriminate between formation models, as is suggested by \citet{Setiawan07, HernanObispo10}.

Unfortunately, observations of young planets are challenging due to their environment. Radial velocity methods lose sensitivity due to the inherent variability of the host star \citep{Saar97} and transit detection and direct imaging methods can be rendered impossible as the planet is obscured by a cloud of dust and gas.
One solution to the observational difficulties posed by young star-disk systems is to search for indirect planet indicators based on interaction with disk dust and gas.

Determining the physical significance of dust structures in transitional and pre-transitional disks is not a new idea. One of the oldest examples of a predicted dust structure is a gap in a protoplanetary disk caused by the direct gravitational influence of a planet \citep{Goldreich80}. The morphology of a gap can be used to infer properties of the disk \citep{Paardekooper04, Fouchet07, Crida06}, with large gaps being indicative of either massive companions or multiple companions \citep{Espaillat14, Dodson-Robinson11, Dong15}. In early studies, due to numerical constraints, it was assumed that the dust and gas in a disk were well mixed, and models of observations still use this method \citep{DAlessio98, Ruge14}. However, due to the imperfect coupling of larger dust grains to the gas, it is argued that the observed structures vary with wavelength \citep{Rice06, Pinilla12, Zhu12, Gonzalez12}. To investigate the effect of imperfect dust-gas coupling, two fluid models coupled with radiative transfer modelling are employed \citep{Pinilla15, Pinilla14, Fouchet10, Zhu12, Owen14}. One important result from two fluid models is the trapping of dust in a planet-induced pressure bump. A pressure bump will reduce the mass of large dust grains interior to the bump; therefore the mm-wavelength signal is reduced such that a cavity is observed. When dust trapping operates in concert with an additional clearing mechanism deeper gaps and cavities are observed \citep{Paardekooper06, Fouchet07, Zhu12}.

The mutual interaction between planets, dust, and gas has also been investigated by the N-body community. The mid to late stages of planet formation are not fully captured by hydrodynamical or N-body approaches alone. As such, combined N-body and Hydro codes are now being used to investigate this epoch as the computational power has only recently become available \citep{Levison12, Levison15, Lambrechts12}. One notable result from the coupling of N-body and Hydro codes is the model of pebble accretion. Pebble accretion relies upon the imperfect (non-negligible, but not dominating) coupling of cm sized particles with gas to efficiently accrete mass onto seed planetesimals \citep{Johansen10, Ormel10, Lambrechts12}. The generation of dust from interactions between planetesimals has been investigated in relation to debris disks \citep[see][]{Wyatt08, Thebault14}, the modelling of giant impacts \citep[e.g.~][]{Kral13}, and the simulation of the mid to late stages of planet formation \citep[e.g.~][]{Leinhardt12, Leinhardt15}. The methods used range from almost purely statistical \citep{Wetherill89, Morbidelli09, Bromley11} to purely N-body \citep{Chambers98, Kokubo00, Raymond09}, with many groups employing a mixture of the two \citep{Spaute91, Weidenschilling97, Bromley11}.

Our first paper, a proof of concept, \citet[hereafter Paper 1]{Dobinson13}, showed a planetary companion can influence the dust distribution of a planetesimal disk. Numerical simulations of gas-free planetesimal disks with an embedded planet of varying eccentricity were conducted with the N-body code, PKDGRAV \citep{Richardson00, Stadel01, Leinhardt09}. The simulations included a planetesimal collision model, RUBBLE, that enabled tracking of growth and disruption of planetesimals and large collision fragments. Collisionally generated second generation dust from planetesimal collisions was modelled in a simple post-processing step. Dust was assumed to sit on an `average' orbit determined from the orbits of both its parent bodies, and did not evolve  (i.e. the orbital parameters and grain sizes could not change after creation). By comparison to an undisturbed control disk, the presence of a planet was shown to have the capability to enhance the visibility of the system and create asymmetries in the dust disk.

The work presented here addresses the main deficiencies of Paper 1 (assumed average orbits and no evolution of dust) by significantly increasing the realism of the numerical scenarios and producing more accurate and useful constraints. The main simulations contain an updated, faster analytical collision model called EDACM \citep{Leinhardt12, Leinhardt15}, and second generation dust is simulated in a more physical manner where both ejection velocity distribution and lifetime are accounted for. In order to increase the realism, non-gravitational forces acting on the second generation dust are also included, such as gas drag from complementary hydrodynamical simulations using the FARGO code \citep{FARGO}, photon pressure, and Poynting Robertson drag. These upgrades provide a more accurate description of the collisional environment present near a planet in a transitional disk. See \S\ref{sec:num_methods} for a more complete discussion. The results of these simulations are presented in \S\ref{sec:results}. First generation primordial dust is excluded from the simulations, however, its effect upon observability of the second generation dust is discussed in \S\ref{sec:discussion}. We give our conclusions in \S\ref{sec:conclusion}.

\section{Numerical Methods}
\label{sec:num_methods}

The numerical techniques used in this investigation can be split into four discrete sections briefly summarised below and discussed in detail in the subsections identified. 

\begin{enumerate}
\item Planetesimal disk (\S \ref{sec:pkdgrav_sims}): $N$-body simulations including particle-particle collisions, a perturbing planet, and inter-particle gravity are used to model the planetesimal population (particles $\ge100$ km).

\item Gas disk (\S \ref{sec:fargo}): Hydrodynamical simulations including an embedded planet are used to provide gas density and velocity maps for fragment simulations. This is needed as collisions in the planetesimal disk produce fragments small enough to be affected by aerodynamic drag.

\item Dust (\S \ref{sec:fse}): Small-sized (10\sups{-2} -- 10\sups{-5} m) collisional debris from the planetesimal disk simulations are integrated directly with additional external forces due to gas and radiation. These fragments would be very small in reality with no significant self-gravity thus they are modelled as test particles that feel the gravitational influence from the star and planet only. Planetesimals are not included.

\item Synthetic Observations (\S \ref{sec:synth_obs}): Dust lifetime, image construction, and radiative transfer modelling are used to create synthetic images and identify observables such as NIR excess, disk asymmetries and gaps that would indicate the presence of an unseen planet. 
\end{enumerate}

The planetesimal and gas disk are treated as separable and numerically modelled independently, however, the results of both are required in order to model the dust and create the synthetic images\footnote{All numerically intensive work was carried out using the computational facilities of the Advanced Computing Research Centre, University of Bristol (http://www.bris.ac.uk/acrc)}.

\subsection{Planetesimal Disk Simulations}
\label{sec:pkdgrav_sims}
The evolution of the planetesimal disk is numerically modelled using a modified version of the parallelised hierarchical tree code, PKDGRAV. This code uses a second order leapfrog integrator, which is symplectic in absence of the gravity tree. The equations of motion for the particles are determined by gravity and physical collisions.

Each simulation begins with $10^6$ equal-mass, 150-km radius planetesimals with a bulk density of $2\textrm{ g cm}^{-3}$ in a circum-stellar disk extending from 0.8 AU to 10 AU around a 1 $M_\odot$ central potential. The planetesimals were initially distributed assuming a minimum mass solar nebula \citep{Weidenschilling77, Hayashi81, Hayashi85} such that the surface density has a standard power-law distribution, $\Sigma(r) = \Sigma_1 r^{-1.5}$, where $\Sigma_1 = 10 \, \mathrm{g}\, \mathrm{cm}^{-2}$ at 1 AU resulting in a total planetesimal mass of $10 \EM$. Eccentricities and inclinations were drawn from a Rayleigh distribution with dispersion $\langle e^2 \rangle = 2\langle i^2 \rangle = 0.001$ \citep{Richardson00}. 

In order to create a more realistic and evolved planetesimal size distribution than the initial uniform distribution, the initial planetesimal disk was integrated using a perfect merging (no fragmentation or erosion) collision model and no embedded planet. This preliminary simulation was integrated until the particle number had reduced by two thirds to $N=3\times 10^5$ (Table \ref{tab:pkdgrav_sims}, row 1). The surface density power-law distribution $\Sigma(r) \propto r^{-1.5}$ is retained, and the size distribution is no-longer single valued (as it was in Paper 1) but is an approximate power law of the form $n(r)dr \propto r^{-3.5}$, with planetesimal radii ranging from the initial 150 km to $\sim$1000 km.

\begin{table}[t]
\tabcolsep=0.11cm
\small
\caption{Planetesimal Simulation Properties}
\begin{tabular}{ccccccc}
\hline
Name & Coll Model& M\subs{pl} & T\subs{grow}(yr) & e\subs{pl} & T\subs{final}(yr) & N\subs{Collisions}\\
\hline \hline
Prelim& Merging \sups{*}& - & - & - & $3.29\times10^4$ & -\\
Control& EDACM& - & - & - & $2\times10^4$ & 9688\\
Ecc0& EDACM & 1\JM & 100 & 0.0 & $2\times10^4$ & 13511\\
Ecc1& EDACM & 1\JM & 100 & 0.1 & $2\times10^4$ & 9498\\
Ecc2& EDACM & 1\JM & 100 & 0.2 & $2\times10^4$ & 11156\\
\hline
\multicolumn{7}{l}{\footnotesize{*This simulation formed the starting point of the others.}}
\label{tab:pkdgrav_sims}
\end{tabular}
\end{table}

After the preliminary phase of perfect merging a planet core is embedded in the planetesimal disk and the collision model is switched to EDACM, which allows multiple collision outcomes. EDACM consists of a set of of scaling laws and collision outcome rules based on data from a series of direct numerical simulations of individual collisions (both $N$-body and hydrodynamical) that characterises the collision type, largest remnants, and fragment size distribution \citep[see][]{Leinhardt12}. EDACM can identify and resolve several collision types including erosion (partial and supercatastrophic), accretion (perfect and partial), and hit-and-run (perfect/bouncing and disruption of the projectile). To reduce computational complexity, these simulations incorporated a size limit. Only fragments larger than 30 km in radius were directly simulated, smaller debris was recorded in ten axisymmetric annular `dust bins' extending from 0.3 to 10.5 AU \citep{Leinhardt05, Leinhardt15}.

In order to incorporate EDACM into PKDGRAV and make the collision model as effective as possible for the tasks presented in this work, modifications were made to EDACM to describe the positions and velocities of the collision fragments as accurately as possible \citep[see][for details]{Leinhardt15}. For self-consistency the modifications to EDACM were derived from the same underlying data used in the development of EDACM.

In this work we completed four main $N$-body planetesimal disk simulations, three with an embedded Jupiter-mass planet and one without to serve as a control case. Given the broad range of exoplanet eccentricities, the eccentricity of the embedded planet was varied (see Table \ref{tab:pkdgrav_sims}).

All disks had the same starting point, the end of the \emph{Prelim} simulation. The embedded planet (when present) was placed at a semi-major axis of 2.8 AU such that the simulation could capture all perturbations from the planet, as eccentricity boosting of the planetesimals can be seen far from the planets orbit. Only one planet is present in the system. Therefore, it is simple to scale the results to any system geometry.

In order to avoid numerical errors from large impulses upon planetesimals near the planet, the embedded planet is grown within the planetesimal disk from 15\EM to 1\JM over 100 yr. Note the time-scale of growth is not physical -- it is a factor of 1000 faster and is used here primarily to create as realistic an initial condition as possible in a practical amount of time (for more detail see Paper 1). 

All planetesimal simulations ignore the gas disk. The mass of gas is only a small fraction of the stellar mass, and any gravity from it can be ignored to the first order \citep{Hartmann98}. In addition, the planetesimals are all initially $>$150 km in radius, thus drag forces upon them are insignificant over the simulation time-scale due to their size. 

Each of the planetesimal simulations was run for $2\times10^{4}$ years, the time-scale could not be too long due to numerical constraints but was made long enough to provide two dust half-lives (see \S\ref{sec:fse}). The time step was set to 0.01 yr, which provided good temporal resolution in all areas of the planetesimal disk.

\subsection{Gas Disk Simulations}
\label{sec:fargo}

Planetesimals are not influenced dynamically by the gaseous component of a circumstellar disk. However, small fragments produced in a collision are. Therefore, hydrodynamical simulations of the gaseous component were performed using the FARGO code for each of the system configurations under investigation.
FARGO is a 2D Eulerian polar grid code that is widely used to model astrophysical disks. These simulations used an initial surface density of $\Sigma(r) = \Sigma_{1g} r^{-0.5}$, with $\Sigma_{1g} = 1780 \,\mathrm{g}\, \mathrm{cm}^{-2}$, an aspect ratio of 0.05, and an $\alpha$-viscosity of $2\times10^{-3}$, similar to the values used by \cite{Zhu12}. To avoid edge effects the FARGO simulations extend beyond the dimensions of the planetesimal disk from 0.4 to 50 AU. A 1 \JM planet is positioned at 2.8 AU (co-incident with the planetesimal disk simulations). The system is integrated for $1\times 10^4$ yrs until it reaches steady state.

Note that the planetesimal disk is assigned a different surface density profile than the gas disk. This is because, as we are assuming the core-accretion model is correct, the growth time-scale of the planetesimals scales with semi-major axis, resulting in faster growth in the inner regions of the disk when compared to the outer regions \citep{Paardekooper10}. Thus, the surface density profile of the $\sim$100 km planetesimals should be more centrally peaked than the gas profile. To reflect this, we have used the minimum-mass solar nebular model for the planetesimal surface density, giving $\Sigma(r) \propto r^{-1.5}$ \citep{Hayashi81}, and have assumed a constant aspect ratio \citep{Takami14} and accretion rate for the gas disk which leads to the $\Sigma(r) = \Sigma_{1g} r^{-0.5}$ surface density profile \citep{Zhu12}.

\subsection{Dust Model}
\label{sec:fse}

The planetesimal disk simulations can cope with erosion and fragmentation of a planetesimal. However, by necessity a size limit was imposed upon the simulated particles. Therefore, small collision fragments were simulated in a secondary code, the fragment simulation engine (FSE).
FSE takes the collisions from a planetesimal disk simulation, models the trajectories of the small fragments using a modified version of the EDACM model \citep{Leinhardt15}, and simulates their orbits.

A second order leapfrog integrator similar to the one used in PKDGRAV with a time step of 0.01 years. Technically, any size of fragment can be simulated, but in this work we restrict ourselves to small fragments as inter-fragment gravity is not included. Fragments of size $10^{-3}$ m to $10^{-5}$ m are simulated as it is mm to $\mu$m-sized particles that have the largest effect upon visibility in the radio and infra-red, wavelengths in which transitional disks are typically observed (see \S \ref{sec:synth_obs} for more details). 

Small particles simulated by FSE are affected by gas drag. Therefore, the gas disk simulations (\S \ref{sec:fargo}) provide the gas properties at all positions in the simulated protoplanetary disk. In the planetesimal disk simulations the planetesimals are large enough that to first order the direct effects of gas can be ignored but inter-particle gravity cannot. When modelling the dust the opposite is true, namely, the dust particles are small so aerodynamic drag from the gaseous accretion disk will affect the orbits of the smallest collision fragments. This is incorporated into the force calculations by the drag equation
\begin{equation}
	\mathbf{F}_\textrm{d}=\frac{- \rho_g (\Delta \mathbf{v})^2 A C_\textrm{d}}{2},
	\label{eq:gas_drag}
\end{equation}
where $\mathbf{F}_\textrm{d}$ is the drag force, $\Delta \mathbf{v}$ is the difference in fragment and gas velocities, $\rho_g$ is the gas density, $A$ is the cross-sectional area of a fragment in the direction of travel (we assume a sphere), and $C_\textrm{d}$ is the drag coefficient which varies depending upon the drag regime \citep{Weidenschilling77} i.e.
	\[C_\tr{d} = \frac{8}{3}\frac{\bar{v}}{v} \quad\tr{in the Epstein regime},\]
	\[C_\tr{d} = 24 \tr{Re}^{-1} \quad \tr{for Re $<$1},\]
	\[C_\tr{d} = 24 \tr{Re}^{-0.6} \quad \tr{for 1$<$Re$<$800},\]
	\[C_\tr{d} = 0.47 \quad \tr{Otherwise.}\]
Where $\bar{v}$ is the thermal velocity of the gas, Re denotes the Reynolds number of the flow, and the Epstein regime is characterised by the dust radius, $a < 9/4\lambda$, where $\lambda$ is the mean free path of a molecule.

Due to the size range of material simulated it is necessary to include photon pressure and Poynting-Robertson drag in addition to the aerodynamic drag from the gas disk. The photon pressure is included as an additional force following \citet{Nichols1903}
\begin{equation}
	\mathbf{F}_{phot} = \beta F_{G} (1 - \mathbf{v} \cdot \hat{\mathbf{r}})\hat{\mathbf{r}},
\end{equation}
where
\begin{equation}
	\beta = \frac{3 L_{*}}{16 \pi G M_{*} c a \rho}
\end{equation}
is the ratio between radiation forces and gravitational forces for a given particle, $L_{*}$ is the stellar luminosity, $G$ the gravitational constant, $M_{*}$ the stellar mass, $a$ the radius of the particle, $\rho$ the density of a particle, $c$ is the speed of light, $F_{G}$ the gravitational force of the star, $\mathbf{v}$ is the velocity of a particle, and $\hat{\mathbf{r}}$ is the unit radial vector from star to particle.

Poynting-Robertson drag \citep{Robertson37, Burns79} is included in FSE as
\begin{equation}
	\mathbf{F}_{PR} = \beta F_{G} (-\mathbf{v}),
\end{equation}
where the symbols have the same definition as above. 

In FSE, fragments are modelled as test particles which do not feel gravity from other test particles. The central star and embedded planet are modelled as gravitating particles (gravity-producing and gravity-feeling), collision detection is not included in the simulation. As test particles have no mutual interaction, FSE lends itself to massive parallelisation and each run was split across 100 cores with each core simulating a different set of collisions that were detected in the main planetesimal simulations (\S \ref{sec:pkdgrav_sims}). Every collision was assigned 100 test particles to follow the velocity field of the fragments found by the modified EDACM model. Each set of collisions was simulated three times (Table \ref{tab:fse_sims}): no gas with small particles (10 $\mathrm{\mu m}$), gas with small particles, and gas with large particles (1 cm).

\begin{table}[t]
\tabcolsep=0.11cm
\small
\centering
\caption{FSE simulation properties}
\begin{tabular}{cccccc}
\hline
\rule{0pt}{2.5ex} 
\# & Parent\sups{*} & Gas & Frag. Size(m) & $N_{collisions}^{\dagger}$ & Figure\\
\hline \hline
1&Control & No & $1\times10^{-5}$ & 9688 & --\\
2&Control & Yes & $1\times10^{-5}$ & 9688& Fig.~\ref{fig:p0_data}\\
3&Control & Yes & $1\times10^{-3}$ & 9688&  --\\
4&Ecc0 & No & $1\times10^{-5}$ & 13511 & --\\
5&Ecc0 & Yes & $1\times10^{-3}$ & 13511& Fig.~\ref{fig:p1_e00_data}\\
6&Ecc0 & Yes & $1\times10^{-3}$ & 13511 & --\\
7&Ecc1 & No & $1\times10^{-5}$ & 9498 & --\\
8&Ecc1 & Yes & $1\times10^{-5}$ & 9498 & Fig.~\ref{fig:p1_e01_data}\\
9&Ecc1 & Yes & $1\times10^{-3}$ & 9498 & --\\
10&Ecc2 & No & $1\times10^{-5}$ & 11156 & --\\
11&Ecc2 & Yes & $1\times10^{-5}$ & 11156 & Fig.~\ref{fig:p1_e02_data}\\
12&Ecc2 & Yes & $1\times10^{-3}$ & 11156 & --\\
\hline
\multicolumn{5}{l}{\footnotesize{* The planetesimal simulation we use the collisions of.}}\\
\multicolumn{5}{l}{\footnotesize{$\dagger$ The number of collisions in the parent simulation.}}\\
\label{tab:fse_sims}
\end{tabular}
\end{table}

\subsection{Synthetic Observations}
\label{sec:synth_obs}

The Fragment Image Reconstruction Engine (FIRE) creates images from FSE, PKDGRAV and FARGO output files, and also creates RADMC3D input files. FSE files provide the dust density information for maps and RADMC3D input, FARGO files provide gas density maps, and PKDGRAV files ensure alignment between different maps. Dust lifetime algorithms are applied to collision fragments. 

The visibility of dust is directly related to its size. FSE simulates dust grains in a specific size range, and outside that size range any material is assumed to be non-visible. The size of dust can be changed by two main pathways: fragmentation to a smaller size, or coagulation to a larger size. Note, physical removal of dust via PR-drag, photon pressure, and gas-drag is accounted for in the FSE. The exact balance between fragmentation and coagulation of dust is unknown and is an ongoing area of research but makes a large difference in the `visibility lifetime' of the dust. If dust never changes size then growth to protoplanets would be impossible, whereas if dust rapidly changes size the observable signatures of protoplanetary disks would quickly dissipate. To model the change in dust size via fragmentation and coagulation we assume the `visibility lifetime' can be treated as an exponential decay.

The `visible' mass, dependent upon a decay constant is varied as
\begin{equation}
P_\textrm{viz}(t) = \mathrm{e}^{(T_\textrm{\tiny creation} - t)/\lambda},
\end{equation}
where $P_\textrm{viz}(t)$ is the fraction of visible dust mass at simulation time $t$, $T_\textrm{\footnotesize creation}$ is the creation time of the particle, $t>T_\textrm{\footnotesize creation}$, and $\lambda$ is e-folding timescale. This represents a certain proportion of available grains becoming non-visible in some way \citep{Dullemond05}. In this work $\lambda = 10^4/\mathrm{ln}(2)$ yr, resulting in a half-life of $10^4$ yr \citep{Adams04}. 

The mass of dust from each collision was found by assuming a power-law distribution of $n(r)dr\,= \eta\,r^{-3.5}dr$ (where $\eta$ is a normalising factor), which gives the mass of dust between the sizes $r_1=1\times10^{-3}$ m and $r_2=1\times10^{-5}$ m as
\begin{equation}
	M(r_1, r_2) = \eta'(r_1^{0.5} - r_2^{0.5}),
	\label{eq:dust_mass}
\end{equation}
where $r_1 > r_2$, and $\eta'$ is a different normalising factor found via $\eta' = M_{rem} M_{slr}^{-0.5}$, where $M_{slr}$ is the mass of the second largest remnant computed via the EDACM code, and $M_{rem}\,=\,M_{total}-(M_{lr}+M_{slr})$ is the mass that would become debris (i.e. mass not included in the largest and second largest remnants of a collision). Therefore, net erosive collisions will contribute more to the mass of dust in a system than net growth collisions. The mass found via (eq. \ref{eq:dust_mass}) scales the mass of the tracer particles.

From the dust density, synthetic images were obtained using the radiative transfer code RADMC3D.  Small dust was modelled as two dust species, amorphous carbon and silicates, at $0.1 \mu m$ and $0.631 \mu m$, each species has relative abundances of 0.2 and 0.8 respectively, in line with interstellar dust \citep[\S12.4.1]{Kruegel03}. Larger dust from $1 \mu m$ to $1000 \mu m$ is modelled using simple Mie scattering spheres with three size bins per decade.

Dust of all sizes should be created in a collision. However, the smallest size is approximately defined by the dust blow-out radius, and the largest is when emission at IR wavelengths (used for observing exozodis) and sub-mm wavelengths (which transitional disks are typically observed in) is no longer significant.

\section{Results}
\label{sec:results}

\begin{figure*}
	\centering
	\includegraphics[width=1.0\textwidth]{\pic/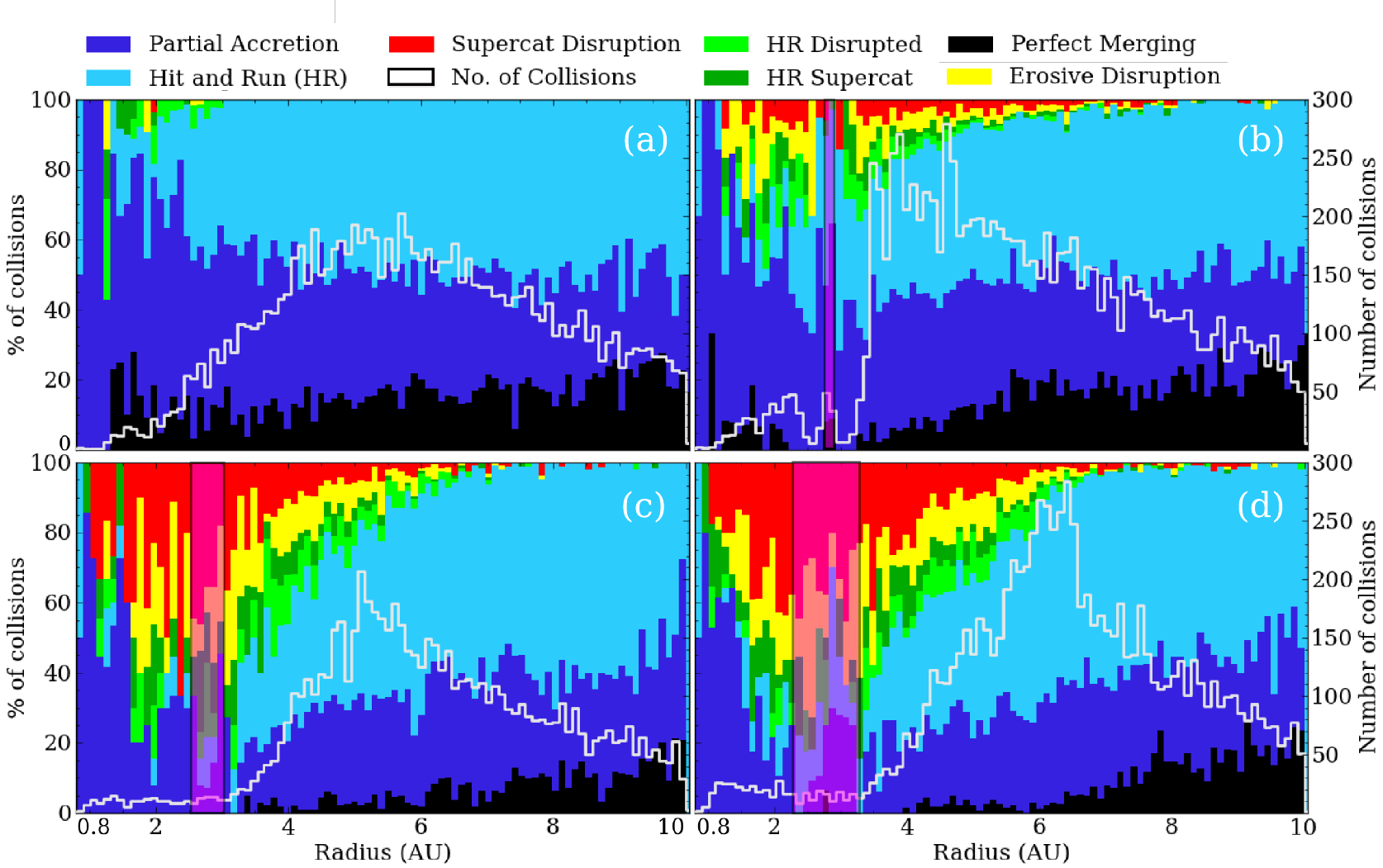}
	\caption{\label{fig:all_collprops}Fraction of collision type with radius for the Control simulation (a), and main simulations Ecc0 (b), Ecc1 (c), Ecc2 (d). Collision type is indicated by the colour key above (a, b), the height of the stacked bars indicates what fraction of the total number of collisions are which type (right hand axis). The white line shows the number of collisions in each bin (right hand axis). The planet is situated at 2.8 AU, and has an eccentricity of 0.0, 0.1, and 0.2 in frames (b), (c), and (d) respectively, the orbital range of the planet is depicted with the magenta area.}
\end{figure*}

\begin{figure*}
	\centering
	\includegraphics[width=1.0\textwidth]{\pic/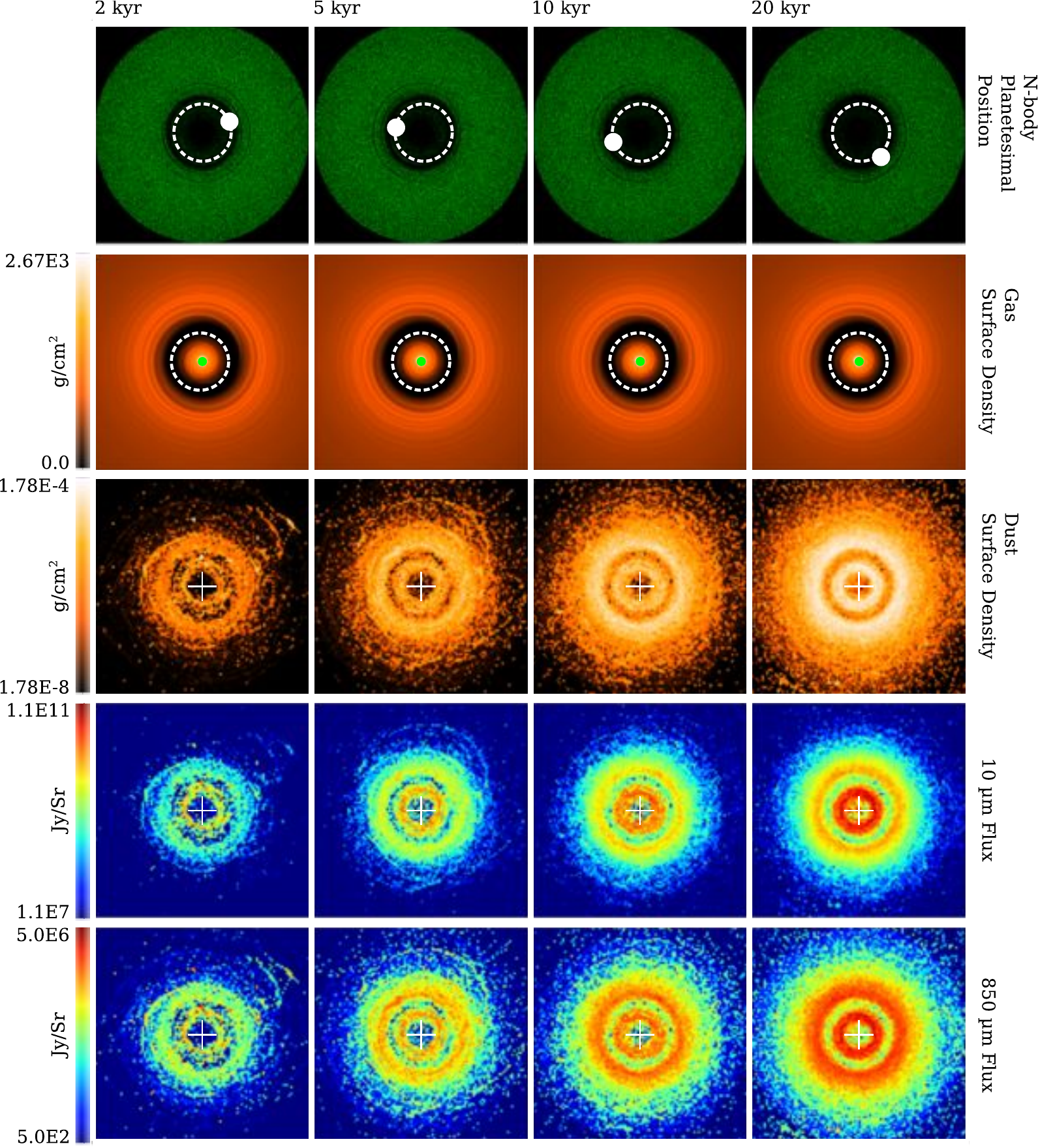}
	\caption{\label{fig:p1_e00_data} Summary of output from simulation 5. From top to bottom: Positions of planetesimals over time, Gas surface density from supporting FARGO simulations (green areas are outside the grid), Dust surface density when lifetime is modelled as an exponential decay (decay constant of $10^4$ yr), Flux from dust (no stellar flux included) from radiative transfer modelling using RADMC3D at 10 $\mathrm{\mu m}$, and at 850 $\mathrm{\mu m}$. White dotted line indicates approximate planet orbit, white circle indicates approximate planet position, white cross indicates barycentre of system.}
\end{figure*}

\begin{figure*}
	\centering
	\includegraphics[width=1.0\textwidth]{\pic/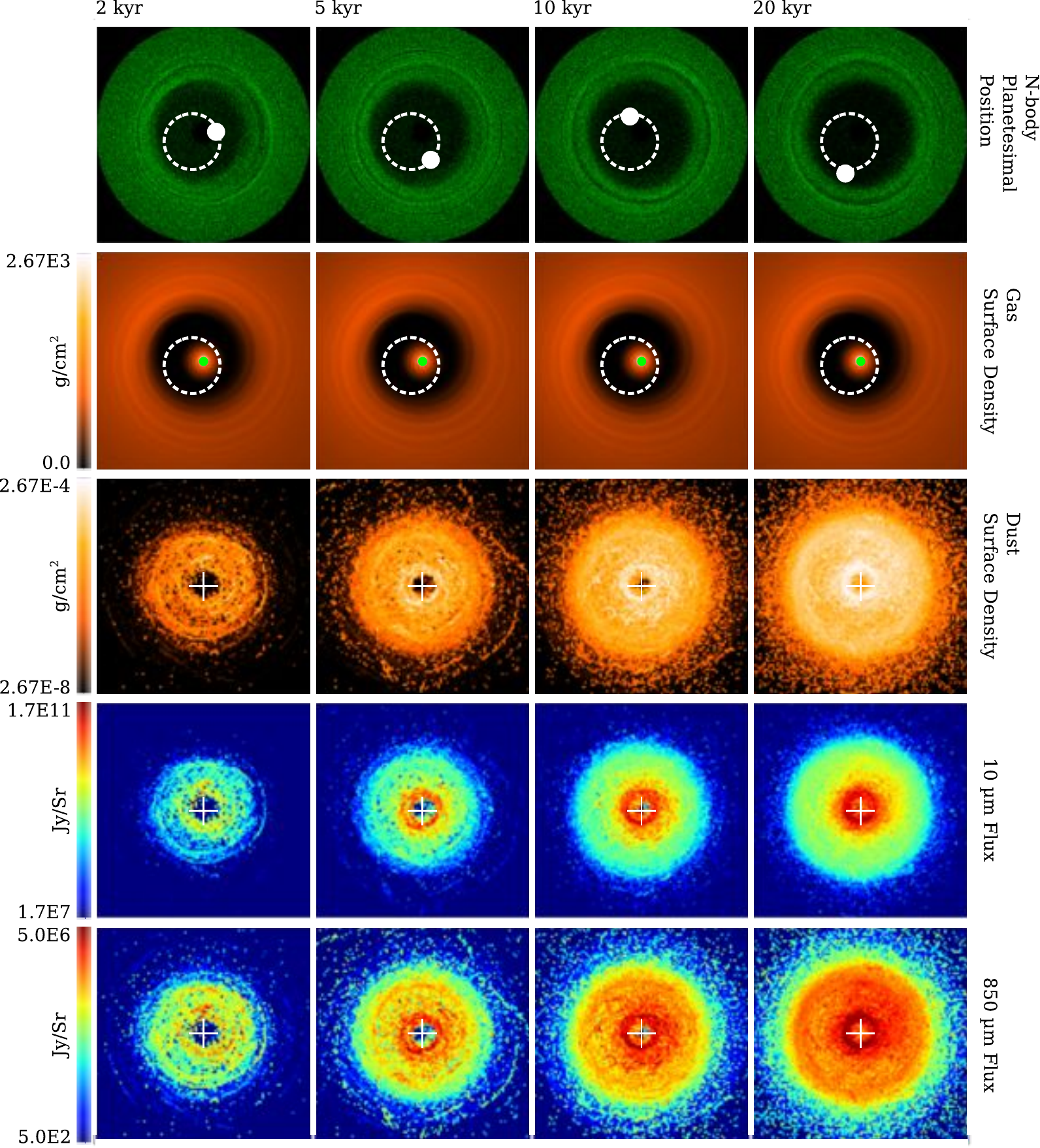}
	\caption{\label{fig:p1_e02_data} Summary of output from simulation 11. From top to bottom: Positions of planetesimals over time, Gas surface density from supporting FARGO simulations (green areas are outside the grid), Dust surface density when lifetime is modelled as an exponential decay (decay constant of $10^4$ yr), Flux from dust (no stellar flux included) from radiative transfer modelling using RADMC3D at 10 $\mathrm{\mu m}$, and at 850 $\mathrm{\mu m}$. White dotted line indicates approximate planet orbit, white circle indicates approximate planet position, white cross indicates barycentre of system.}
\end{figure*}

\begin{figure*}
	\centering
	\includegraphics[width=1.0\textwidth]{\pic/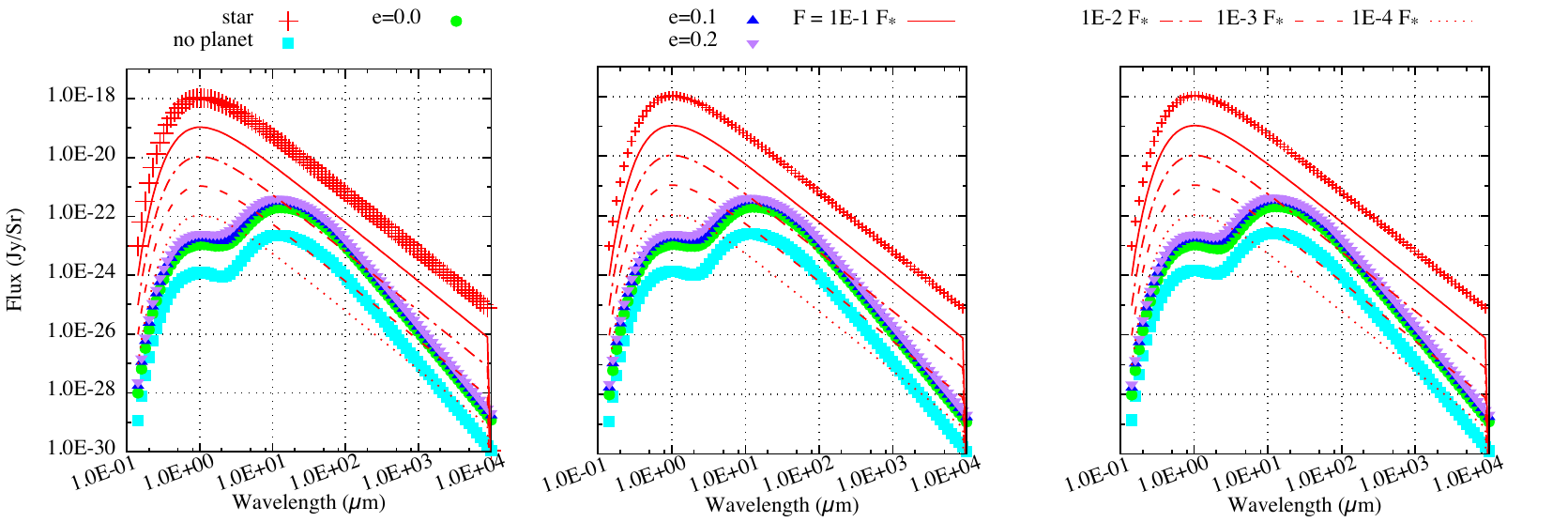}
	\caption{\label{fig:spectrum}Spectrum of dust surface density with exponential decay applied ($10^4$ yr timescale). From left to right: no gas component, gas with small ($1\times10^{-5} \mathrm{m}$) fragments, gas with large ($1 \times 10^{-3} \mathrm{m}$) fragments. Red `+'s are the stellar flux, other markers are different simulations. Lines (solid, dot-dashed, dashed, dotted) are guide lines showing different fractions of stellar flux. All frames are taken at 20 kyr, with a 10 kyr decay constant and computed with RADMC3D.}
\end{figure*}

Figure \ref{fig:all_collprops} shows the collision type versus disk radius at the end state of the four main simulations. Collisions were summed over time and binned as a function of semi-major axis for the Control, Ecc0, Ecc1, and Ecc2 simulations (Table \ref{tab:pkdgrav_sims}), each of which provides collision data for three FSE simulations (Table \ref{tab:fse_sims}). Colours denote the type of collision, with proportions shown by stacked bars relating to the left hand y-axis, the white line shows number of collisions in a given bin relating to the right hand y-axis. Collision types are as follows: perfect merging -- colliders merge inelastically and no debris is produced, partial accretion -- debris is produced but there is net growth of one collider, erosive disruption -- debris is produced and one or both colliders are smaller, supercatastrophic disruption -- debris is produced and neither collider survives,  hit-and-run -- colliders bounce without changing mass, hit-and-run disrupted -- the smaller collider is eroded and produces dust while the larger collider is unaffected, hit-and-run supercatastrophically disrupted -- the smaller collider is destroyed and produces dust while the larger collider is unaffected. The collisions shown in Fig.~\ref{fig:all_collprops} are only planetesimal-planetesimal collisions and do not include collisions with the $1 M_J$ planet, collisions with the planet are purely accretive and do not produce any debris. 

The total number of collisions (white line, right hand scale) is approximately equal for each simulation (see Table \ref{tab:fse_sims}). However, the radial distribution is very different. For simulations including a planet (b, c, d) the region near the planet at 2.8 AU has fewer collisions than the control case (a). The peak in the number of collisions is comparable or larger in number. The reduction near the planet coincides with an increase in destructive collisions and the peak is shifted to a larger radius where there are fewer destructive collisions. We suggest the reduction in collision number is due to stirring from a planet causing highly destructive collisions in its vicinity, such that planetesimals experience few collisions before destruction. The shifting of peak collisions is also due to stirring from the planet, but of a lower magnitude. Planetesimals are disturbed so that they collide with increased frequency but with low enough velocities that they survive the encounters.

Erosive collisions (greens, yellow, and red) become more frequent throughout the radial extent of the disk when the planetary perturber is introduced, and increase in frequency with eccentricity. Also, the proportion of non-erosive collisions (blues and blacks) decreases in the same manner, or remains steady. The increase in erosive collisions due to the presence of a planet is attributed to the higher level of gravitational stirring, which pumps up the mutual velocities of the planetesimals increasing the number of energetic collisions, this also explains the decrease in non-erosive collisions as these are generally lower energy. The extra erosive collisions caused by a more eccentric planet are due to the same mechanism, but in a more extreme way. The planet can influence more planetesimals and put them onto more eccentric orbits, which increases the collision energies more than a non-eccentric planet.

Figures \ref{fig:p1_e00_data} and \ref{fig:p1_e02_data} show the main results from simulations 5 and 11. From the top the rows are: planetesimal position from PKDGRAV simulations, gas surface density from FARGO simulations, dust surface density from the FSE secondary simulations, 10 $\mu$m flux, and 850 $\mu$m flux computed with the RADMC3D radiative transfer code. The columns increase in time from left to right. Similar figures for simulations 2 and 8 can be found in the appendix (Fig.~\ref{fig:p0_data}, \ref{fig:p1_e01_data}).

In Figures \ref{fig:p1_e00_data} and \ref{fig:p1_e02_data} the planetesimal position and gas density show the clearing of a gap co-incident with the planet orbit. As expected, the gap is larger but shallower in the eccentric case. At later times the gap becomes cleaner due to particle growth and collisional destruction which removes material, this is more noticeable in the eccentric case.

However, the dust surface density frames do not show as much clearing as the planetesimal position and gas density frames. The non-eccentric case (sim.~5, Fig.~\ref{fig:p1_e00_data}) has a narrow ring of cleared space, whereas the eccentric case (sim.~11, Fig.~\ref{fig:p1_e02_data}) has two brighter inner and outer rings with a lower surface density between them. In both cases there are two brightness peaks, one interior and one exterior to the planet's orbit. This is similar to the structures found in Paper 1. However, disk asymmetry is less pronounced in this work as these simulations account for the eccentricity distribution of the second generation dust rather than assuming a single orbit.

The flux density frames, both 10 $\mu$m and 850 $\mu$m show an increase in peak flux between the non-eccentric (sim.~5) and eccentric (sim.~11) case. This is due to the previously mentioned dust rings interior and exterior to the planetary orbit. In the eccentric case (sim.~11) the inner ring is closer to the star and therefore hotter, thus contributing more to the flux. Additionally, the eccentric planet forces planetesimals onto eccentric orbits, creating more dust (via more erosive collisions) and putting fragments on eccentric orbits which increases the dust mass close to the star.

Figure \ref{fig:spectrum} shows a spectral energy distribution (SED) of the emission from the disk compared to the emission from the star. The red `+' signs show the stellar emission, blue squares show \emph{Control} simulation data, green dots show \emph{Ecc0} data, upright blue triangles show \emph{Ecc1} data, inverted purple triangles show \emph{Ecc2} data, and red lines (solid, dotted, dashed, and dot-dashed) show decreasing fractions of the stellar flux. 

Emission enhancement due to second generation dust caused by the presence of the planet is easily seen. The peak emission from the no-planet case (sim.~2) is approximately 1/10th of the planetary emission (sims 5, 8, 11). The SEDs for the planetary simulations (dots and triangles) are so similar as to be practically indistinguishable even between the gaseous and non-gaseous cases. However, some trends between the planetary cases can be observed. As eccentricity increases flux increases for the left and right frames. The central frame has less distinction between eccentricities due to the small particle sizes being entrained in the gas disk. Observationally, the enhanced flux due to second generation dust would be noticeable using nulling interferometry if collisional material was the only dust source present, but the \emph{Control} case would not (see \S \ref{sec:discussion}).

\section{Discussion}
\label{sec:discussion}

So far we have considered the case where second generation dust is the only source of dust present in a transitional disk. This may not be the case if they retain some of their first generation primordial dust, albeit at a lower surface density, than their protoplanetary counterparts.

\begin{figure*}
	\centering
	\includegraphics[width=1.0\textwidth]{\pic/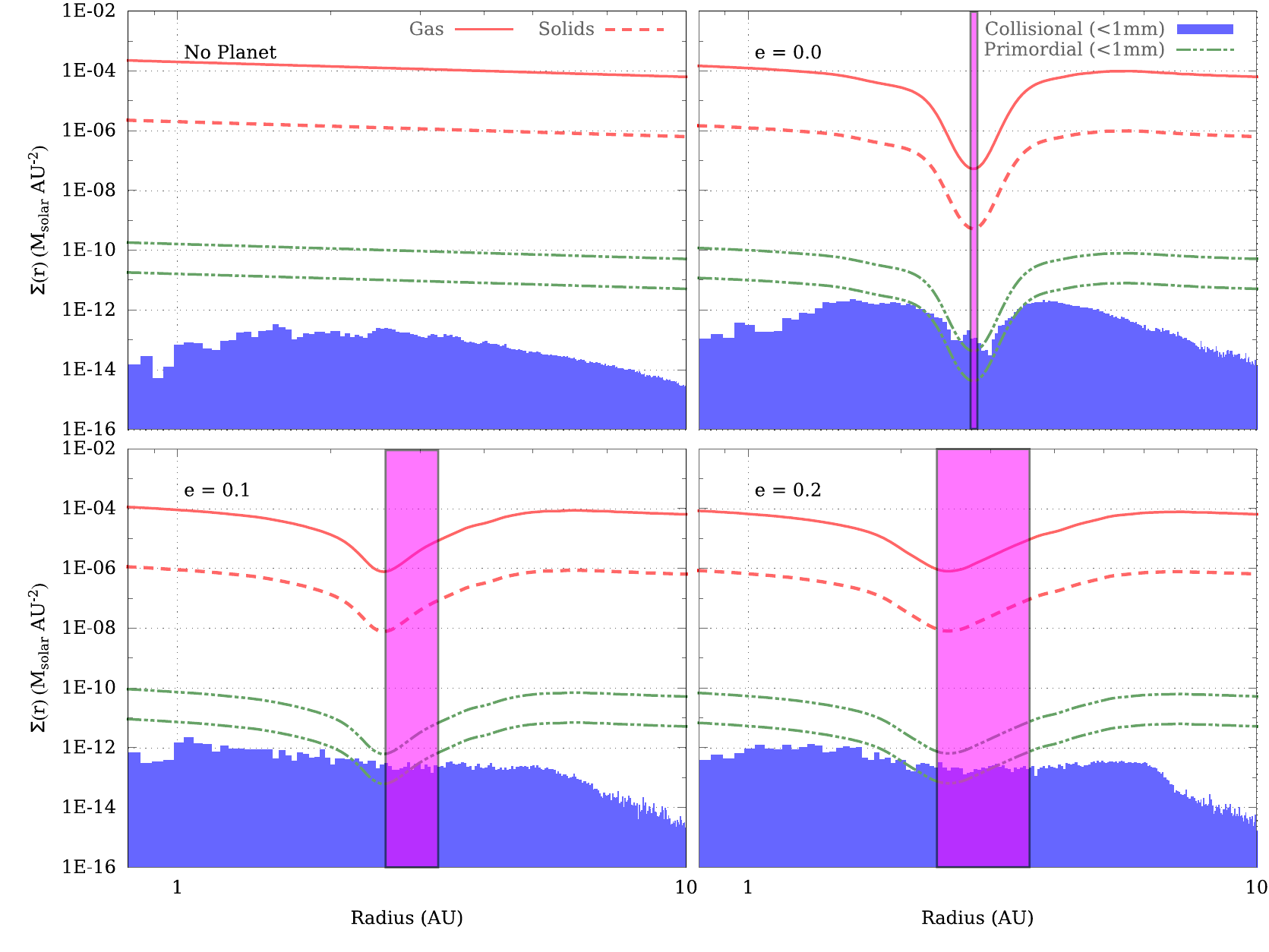}
	\caption{\label{fig:primordial} Dust surface density against radius for each of the four main simulations. Solid red line is the gas surface density, dotted red line is the solid material assuming a 100:1 gas to solids ratio, green dot-dashed line (with fill pattern) is an estimate for the surface density of small ($<1$ mm) dust, blue bars are the collisional material from the indicated simulations. All frames are taken at the end of the simulations. The orbital extent of the planet is depicted with the magenta rectangle, with inner and outer edges corresponding to periastron and apoastron respectively.}
\end{figure*}

Figure \ref{fig:primordial} shows the surface density of simulated second generation dust against the implied first generation material present in the gas simulations. The solid red line shows the gas density, the dotted red line shows surface density of solid material assuming the usual 100:1 gas:solids ratio, green shaded area shows an estimate for small ($<$1 mm) first generation dust, and blue bars show the second generation dust from the simulations.

The estimate for small first generation dust grains used the same power-law assumption as the size of collisionally generated material summarised in (eq.~\ref{eq:dust_mass}) such that
\begin{equation}
	M_{<\mathrm{1\,mm}} = M_\mathrm{solid}(r_1^{0.5} - r_2^{0.5})/(r_{cut}^{0.5} - r_2^{0.5})
	\label{eq:pri_dust_mass}
\end{equation}
where $M_\mathrm{solid}$ is the mass of solid material assuming a 100:1 gas:solid mass ratio, $r_1 = 1\times10^{-3}$ m is the radii of the largest observable first generation dust, $r_2 = 1\times10^{-5}$ m is the smallest first generation dust, $r_{cut} = 1\times10^{5}$ m is the cut off below which we do not continuously simulate objects. In this case, our starting planetesimal size.

The upper limit of the green region, which shows an estimate of the mass of small ($<$1 mm) first generation dust, is found by applying (eq.~\ref{eq:pri_dust_mass}) to the estimate of solid material (red dotted line). The lower limit is found by assuming a lowering of gas surface density over the lifetime of a disk by an order of magnitude \citep{Jones12}.

Second generation dust from the no-planet case (sim.~2) would be rendered invisible by the large mass of first generation material (and any first generation material would be undisturbed). However, assuming a conservative estimate for the primordial material, the planetary cases (sims 5, 8, 11) see a comparable mass of first generation to second generation dust throughout the disk, and within the gaps second generation dust is the dominant source. The zero eccentricity case (sim.~5) would be the most obvious, with bright rings either side of the planet and a possible enhancement coincident with it. For more eccentric planets other markers such as asymmetries and dust structures would need to be relied upon.

The mass of first generation dust estimated from (eq.~\ref{eq:pri_dust_mass}) assumes that the size distribution of growing dust and pebbles follows the same power law as planetesimals, and that dust coagulation leads to the formation of planetesimals. The well known cm and m barriers \citep{Weidenschilling77, Morbidelli08} along with the fragility of 1 km planetesimals \citep{Leinhardt09, Nelson10} will perturb this power law, especially in the case of the cm barrier which can result in mass `piling up' in smaller sizes. Also, if planetesimals are not constructed via dust coagulation, then gravitational instability formation pathways such as turbulent concentration and streaming instability \citep[e.g.][]{Johansen06, Cuzzi08, Bai10, Gressel11} can circumvent the growth of intermediate sized objects completely such that (eq.~\ref{eq:pri_dust_mass}) no longer holds, even as an approximation.

Dust production (and therefore mass) is enhanced by the presence of a planetary body by a factor of $\sim$10 (Fig.~\ref{fig:primordial}).  Taken in combination with Fig.~\ref{fig:all_collprops} we conclude that in the no-planet case `Partial Accretion' (dark blue) collisions provide the main source of dust, whereas in the planetary case `Supercatastrophic' (red and dark green) collisions provide the main source and are much more efficient at dust production, as would be expected. For our scenario, the disk emission (Fig.~\ref{fig:spectrum}) is 1/100th of the stellar emission at peak ($F_{disk}/F_* \sim 1/100$) this is easily detectable by nulling interferometery which can claim detections of exozodis with $F_{disk}/F_* \geq 10^{-4}$ in the N-band \citep{Millan-Gabet11}, and could therefore also detect the no-planet cases. The survey conducted in \citet{Absil13} was sensitive to $F_{disk}/F_* \geq 10^{-2}$ in the K-band which would render our model systems undetectable. However, our simulated disk was truncated at 0.8 AU due to numerical reasons, if this restriction was relaxed the planetary cases may be marginally detectable. If a sufficient planetesimal population survives to late times, gravitational stirring is a possible sources of young exozodis.

As this simulation only spanned 20,000 yr the long-term observability of these systems is unknown. Due to gravitational stirring, even a reduction of collision rate may keep the planetary cases observable due to the higher proportion of violently erosive collisions, which are the main generator of dust, and an overall decrease in primordial dust mass over time.

The planetesimals simulated are of the order $\sim$100 km, which we have assumed are not affected significantly by aerodynamic drag. On the timescales of our simulations this is justified, however, eccentric planetesimals can become circularised on $\sim$1 Myr timescales and migrate on $\sim$5 Myr timescales \citep{Grishin15}. Far from the perturbing planet this will cause a decrease in the planetesimal collision rate, however, when the perturbations from the planet are strong (which is the case in this work) the authors argue the planetesimals will retain their eccentric orbits. The long term behaviour of eccentric planetesimals in the vicinity of a planet and under the influence of aerodynamic drag should be investigated in future work as the long term observability of a system is tied to the collision frequency.

Interestingly, the presence of gas in the simulations does not change the spectral energy distribution substantially (Fig.~\ref{fig:spectrum}). This may be caused by fragments being stirred by the planet on a much shorter time scale than the gas entrainment time scale. Or possibly the gas near the planet is perturbed in the same manner as the fragments, such that entrainment does happen but is qualitatively identical to the no gas case. However, the presence of gas does `wash out' any non-axisymmetric structures in the eccentric simulations.

In the earlier phases of the simulation bright spots and arcs of dust are easily visible in the dust surface density plots. As time moves on these structures, while still present, are more difficult to observe. In all cases, these are due to collisions that occurred just prior to the frame in question and the fragments have not had sufficient time to spread over their orbit. Possibly this is analogous to bright spots seen in debris disks recorded by \citet{Su15}.

The double ring structure of a zero eccentricity planet would be detectable with ALMA if gas is depleted by an order of magnitude with respect to a protoplanetary disk. Providing a sensitivity of 10 $\mu$Jy with a five hour observation under ideal circumstances \citep{ALMAcalc}.

\section{Conclusion}
\label{sec:conclusion}

In this work we present results from 4 simulations (Table \ref{tab:pkdgrav_sims}), each of which has had its second generation dust modelled in three separate ways (Table \ref{tab:fse_sims}). All simulations were integrated with the PKDGRAV code using the EDACM collision model. During these simulations each collision was recorded, and a modified version of the EDACM collision model was used to produce small dust fragments with accurate velocities. The orbits of the dust fragments were integrated with a simple N-body code using the kick-drift version of the leapfrog integrator and split across multiple processors. The resulting dust density was post-processed to account for a $10^4$ yr dust lifetime (measured from the moment of the generating collision) and passed to RADMC3D for radiative transfer modelling.

The presence of gas was simulated by running an analogous FARGO simulation for each PKDGRAV simulation. Once a quasi steady state was reached, the gas density and velocity fields were included in the fragment simulations. The presence of gas had no significant effect upon second generation dust distribution over the the time simulated. However, a reduction, or `wash out', of non-axisymmetric structures is observed when gas is included.

Dust production is enhanced by the presence of a planetary body by a factor of $\sim 10$ (Fig.~\ref{fig:spectrum}). Also, the main source of second generation dust comes from erosive collisions rather than the more common partially accretive collisions. The collisional material has a similar fractional luminosity as exozodis. If a planetesimal population is present in a system at a late time, second generation dust would be detectable with nulling interferometry \citep{Millan-Gabet11, Absil13}.

Second generation dust as a marker for the presence of an unobserved planetary companion would be visible with ALMA. In the case of a circular planet in a system with gas depleted by an order of magnitude with respect to a protoplanetary disk, observations with ALMA would be able to detect a double ring structure (Fig.~\ref{fig:p1_e00_data}).

\section{acknowledgements}
J.D. is grateful for support from NERC, Bristol School of Physics, and McDonald Observatory. Z.M.L. is supported by a STFC Advanced Fellowship. S.D.R. is supported by National Science Foundation CAREER award AST-1055910. N.A.T. is supported by the UK Space Agency/STFC. This work was carried out using the computational facilities of the Advanced Computing Research Centre, University of Bristol. Also, thanks to the anonymous referee whose suggestions have been immensely helpful.

\bibliographystyle{apj}
\bibliography{\thebib/everything.bib}

\appendix
\label{sec:appenndix}
\setcounter{figure}{0} \renewcommand{\thefigure}{A.\arabic{figure}}

\begin{figure*}
	\centering
	\includegraphics[width=1.0\textwidth]{\pic/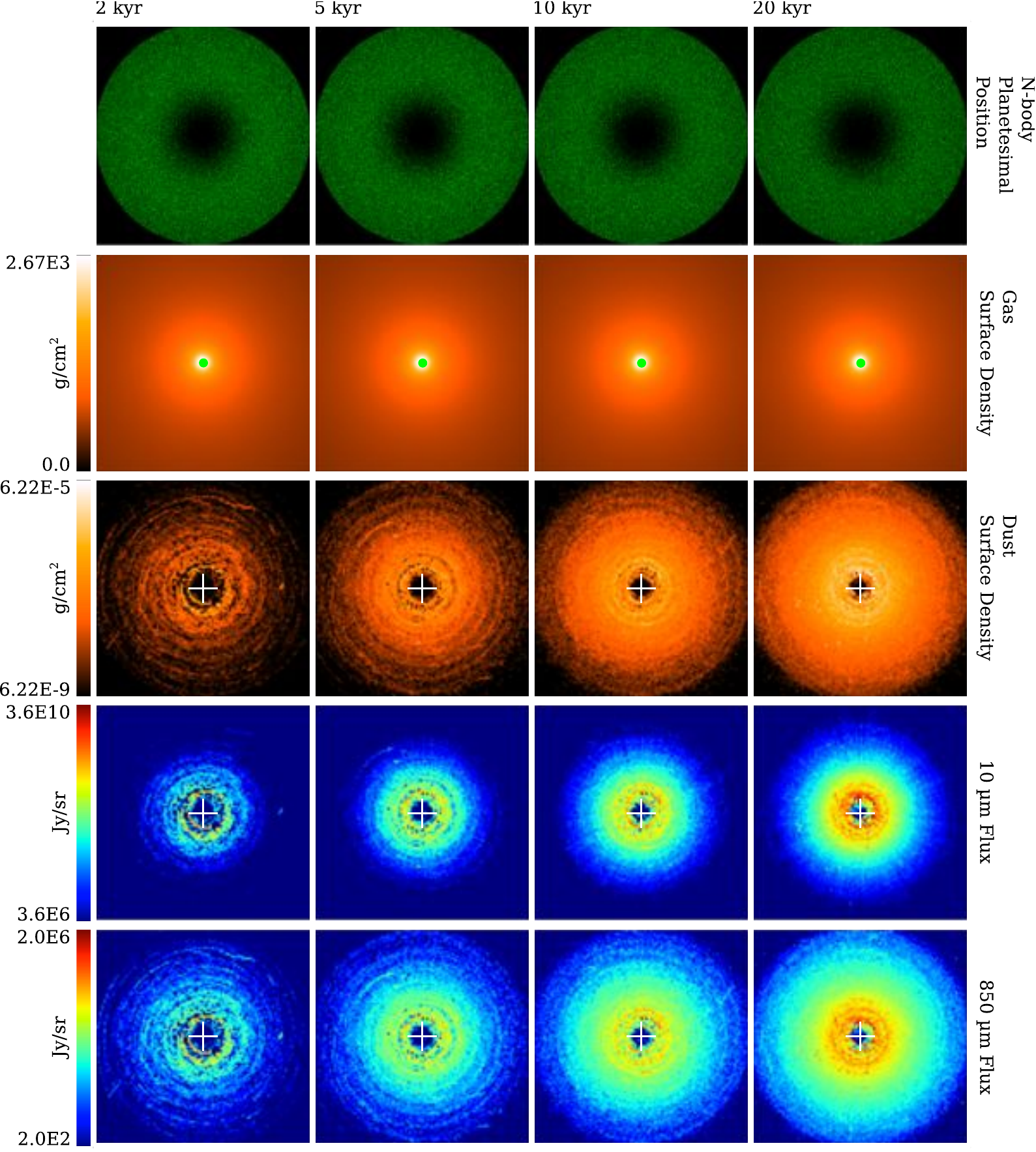}
	\caption{\label{fig:p0_data} Summary of output from simulation 2. From top to bottom: Positions of planetesimals over time, Gas surface density from supporting FARGO simulations (green areas are outside the grid), Dust surface density when lifetime is modelled as an exponential decay (decay constant of $10^4$ yr), Flux from dust (no stellar flux included) from radiative transfer modelling using RADMC3D at 10 $\mathrm{\mu m}$, Flux from dust (no stellar flux included) from radiative transfer modelling using RADMC3D at 850 $\mathrm{\mu m}$. White cross indicates barycentre of system.}
\end{figure*}

\begin{figure*}
	\centering
	\includegraphics[width=1.0\textwidth]{\pic/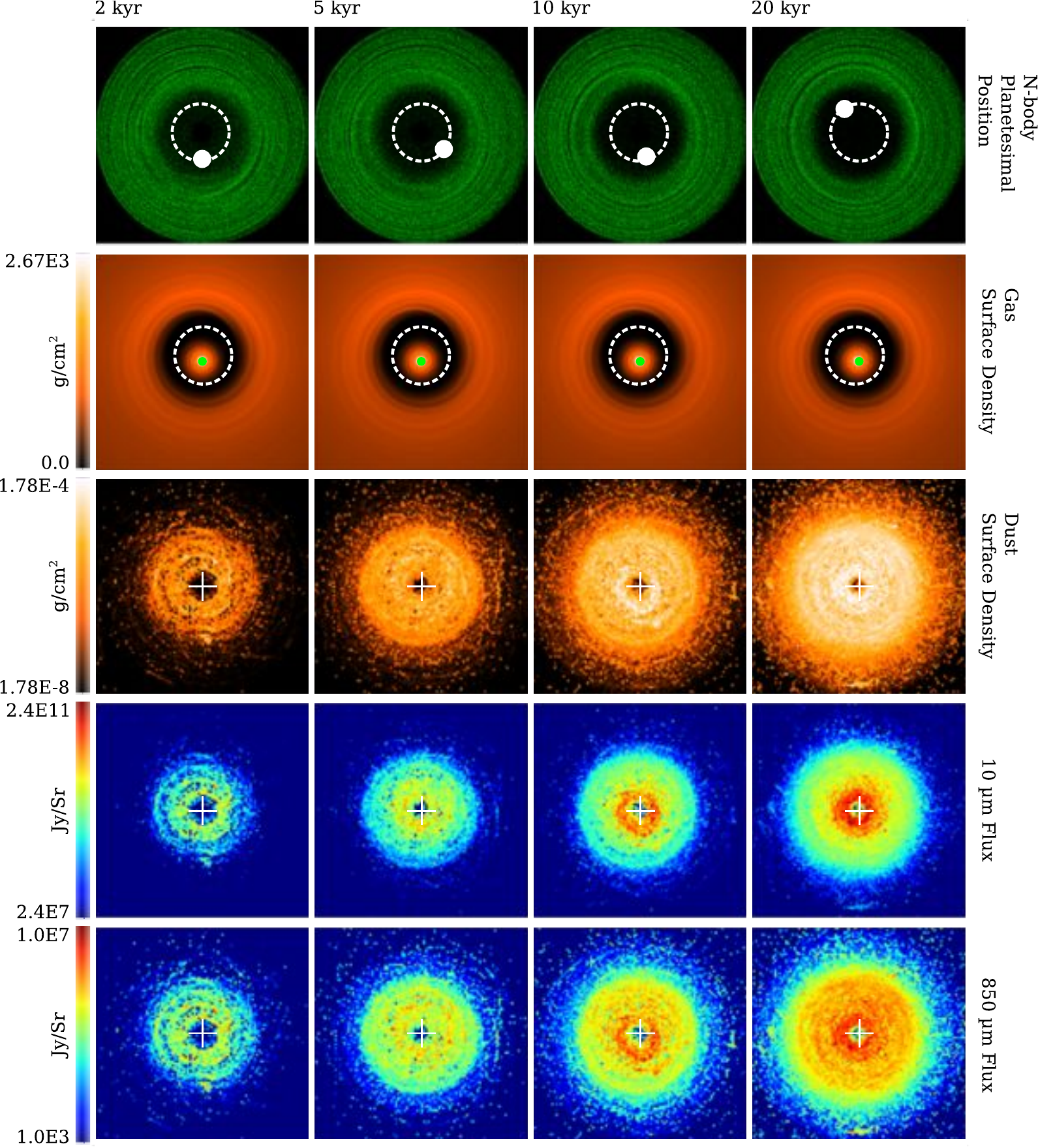}
	\caption{\label{fig:p1_e01_data} Summary of output from simulation 8. From top to bottom: Positions of planetesimals over time, Gas surface density from supporting FARGO simulations (green areas are outside the grid), Dust surface density when lifetime is modelled as an exponential decay (decay constant of $10^4$ yr), Flux from dust (no stellar flux included) from radiative transfer modelling using RADMC3D at 10 $\mathrm{\mu m}$, Flux from dust (no stellar flux included) from radiative transfer modelling using RADMC3D at 850 $\mathrm{\mu m}$. White dotted line indicates approximate planet orbit, white circle indicates approximate planet position, white cross indicates barycentre of system.}
\end{figure*}

\end{document}